\documentclass{article}
\usepackage[utf8]{inputenc}

\usepackage{amsthm}
\usepackage{url}
\usepackage{authblk}
\usepackage{amsmath,amstext,amssymb,amsfonts,verbatim}

\usepackage{color}

\usepackage{graphicx,algorithmic,algorithm,rotating}
\allowdisplaybreaks[1]

\newtheorem{theorem}{Theorem}
\newtheorem{definition}[theorem]{Definition}

\title{\vspace{-1.5cm}Symmetry and Algorithmic Complexity of Polyominoes and Polyhedral Graphs\thanks{An online implementation to estimations of graph complexity is available online at \protect\url{http://www.complexitycalculator.com}}}

\begin{document}

\author{Hector Zenil$^{1,2,3,4}$, Narsis A. Kiani$^{1,2,3,4}$, Jesper Tegn\'er$^{2,3,5}$\\
$^1$ Algorithmic Dynamics Lab, Centre for Molecular Medicine,\\ Karolinska Institute, Stockholm, Sweden\\
$^2$ Unit of Computational Medicine, Department of Medicine,\\Karolinska Institute, Stockholm, Sweden \\
$^3$ Science for Life Laboratory, SciLifeLab, Stockholm, Sweden\\
$^4$ Algorithmic Nature	Group, LABORES	for	the	Natural	and\\Digital Sciences, Paris, France\\
$^5$ Biological and Environmental Sciences and Engineering Division,\\	Computer, Electrical and Mathematical	Sciences and Engineering\\Division,	King Abdullah University of Science and\\	Technology (KAUST), Kingdom	of Saudi Arabia\\
{\{hector.zenil, narsis.kiani, jesper.tegner\}@ki.se}}

\date{}

\maketitle

\begin{abstract}
We introduce a definition of algorithmic symmetry able to capture essential aspects of geometric symmetry. We review, study and apply a method for approximating the algorithmic complexity (also known as Kolmogorov-Chaitin complexity) of graphs and networks based on the concept of Algorithmic Probability (AP). AP is a concept (and method) capable of recursively enumerate all properties of computable (causal) nature beyond statistical regularities. We explore the connections of algorithmic complexity---both 
theoretical and numerical---
with geometric properties mainly symmetry and topology from an (algorithmic) information-theoretic perspective. We show that approximations to algorithmic complexity by lossless compression and an Algorithmic Probability-based method can characterize properties of polyominoes, polytopes, regular and quasi-regular polyhedra as well as polyhedral networks, thereby demonstrating its profiling capabilities. \\

\noindent{}\textbf{Keywords:} Kolmogorov-Chaitin complexity; algorithmic probability; algorithmic Coding Theorem; Turing machines; polyominoes; polyhedral networks; molecular complexity; polytopes; information content; Shannon entropy; symmetry breaking; recursive transformation.
\end{abstract}

\section{Connecting information and symmetry}

The literature on the subject of finding connections between symmetry and complexity is sparse, in particular in connection to information theory and algorithmic complexity.

In~\cite{lin}, a relation between symmetry and entropy was suggested in the context of molecular complexity, thereby establishing connections between low symmetry and low entropy or higher (classical---as opposed to algorithmic) information content.


A measure of structural complexity must not be based on classical symmetry. The use of symmetry in the realm of complexity is justified only in the context of combinatorial complexity, establishing equivalences among the diversity of elements of an object with symmetry by way of statistical regularities. An illustrative example is an object such as the mathematical constant $\pi$, which would appear random to an observer uninformed about its deterministic origin, and which has no apparent symmetry even though it is often, if not always, involved in aspects of symmetry related to, e.g., thesphere. The constant $\pi$, however, can be briefly described algorithmically (by a formula or a computer program), and therefore is considered to be of low complexity~\cite{bdmpaper}. It is clear then that classical and algorithmic information theory measure different properties. However, Shannon entropy does not provide a method to have access to the probability distributions thereby heavily relying on ensemble assumptions to which an object in question is supposed to belong. 

There is therefore a need for tools to extract meaningful generalizations from graphs and networks in, for example, computational chemistry. The concept of symmetry has traditionally been important in areas ranging from pure mathematics to biology and molecular complexity in chemistry. In developmental biology, for example, symmetry is a powerful tool to transfer information and build organisms by self-replication communicating information across long distances. In chemistry, symmetries of a molecule and of molecular orbitals forming covalent bonds, have been studied extensively.

In~\cite{toxicologypaper,chemicalnetworks} we explore connections related to structure networks of chemical compounds with toxicological applications. Here, in the current paper, we provide a unified and robust platform for estimating the algorithmic complexity of a graph on the basis of algorithmic information theory that is applicable to both abstract objects---e.g. geometric ones such as polyominoes, polytopes, and graphs. 

We test this notion in terms of simplicity and complexity and in our analysis we find connections between symmetry and complexity. We will then examine properties of symmetry captured by graph descriptions, in order to test and illustrate the algorithmic measure of algorithmic complexity for graphs and networks.

\subsection{Polyominoes}

A polyomino is a collection of cells of equal size that share at least one side. One can think of polyominoes as an extension of dominoes. Polyominoes are familiar because of their use of Tetrominoes (polyominoes of size 4) as introduced in the game of Tetris. The goal of the game of Tetris can actually be defined as the reduction of the Tetromino complexity by tiling, and while solutions to Tetris can easily be achieved by minimization of the spaces between Tetrominoes, the method of algorithmic complexity, by means of finite approximations to algorithmic probability (AP), provides a (numerical as well as practical) alternative to the solution of Tetris other than by the minimization of spaces, with the alternative solution by means of algorithmic complexity producing interesting, efficient packings through minimization of shape complexity by producing tilings of low complexity.



\subsection{Polyhedra, polytopes and polyhedral networks}

\begin{definition}
A polytope is a geometric object with flat sides, and may exist in any general number of dimensions $n$ as an $n$-dimensional polytope or $n$-polytope. 
\end{definition}

For example, a two-dimensional polygon is a 2-polytope and a three-dimensional polyhedron is a 3-polytope.

\begin{definition}
A Platonic solid is a regular, convex polyhedron.
\end{definition}

There are 5 polyhedra that can have the properties of a Platonic solid in 3-dimensions and 13 that are Archimedian.

\begin{definition}
An Archimedean solid is a semi-regular convex polyhedron composed of regular polygons meeting in identical vertices, excluding the 5 Platonic solids and also the prisms and antiprisms.
\end{definition}

\begin{definition}
Every $n$-polytope has a dual structure, obtained by interchanging its vertices for facets, edges for ridges, and so on, generally interchanging its ($j-1$)-dimensional elements for ($n-j$)-dimensional elements (for $j = 1$ to $n-1$), while retaining the connectivity or 
between elements.
\end{definition}

As an example, the tetrahedron is dual to itself, the cube is dual to the octahedron, and icosahedron is dual to dodecahedron. Thus, to find the direct symmetry group of all the 5 Platonic solids it suffices to find the groups of the tetrahedron, cubes, and dodecahedron.

\begin{definition}
An $n$-polyhedral graph (or $c$-net) is a 3-connected simple planar graph on $n$ nodes.
\end{definition}

\subsection{Basics of Graph theory}

\begin{definition}
A graph is an ordered pair $G = (V,E)$ comprising a set $V$ of nodes or vertices and a set $E$ of edges or links, which are 2-element subsets of $V$.
\end{definition}

\begin{definition}
A graph is \textit{planar} if it can be drawn in a plane without graph edges crossing.
\end{definition}

Planarity is an interesting property because only planar graphs have \textit{duals}. 

\begin{definition}
A \emph{dual graph} of a planar graph $G$ is a graph that has a vertex corresponding to each face of $G$, and an edge joining two neighbouring faces for each edge in $G$.
\end{definition}

Dual polyhedra share the same symmetry axes and planes.
 
\begin{definition}
A \textit{uniform polyhedron} is a polyhedron which has regular polygons as faces and is vertex-transitive on its vertices, that is, there is an isometry mapping any vertex onto any other. 
\end{definition}

Every convex polyhedron can be represented in the plane or on the surface of a sphere by a 3-connected planar graph.

In what follows, we will use the terms nodes and vertex, and links and edges, interchangeably.

\section{Methodology}

\subsection{Computability/recursivity}

In computability theory, computable (or Turing-computable) functions are also called recursive functions. Computable functions are the mathematical formalization of the intuitive notion of an algorithm. 

\begin{definition}
A function $f:\mathbb {N}^{k}\rightarrow \mathbb{N}$ is computable if and only if there is a Turing machine that, given any $k$-tuple $x$ of natural numbers, will produce the value $f(x)$.
\end{definition}

\noindent In other words, a function is computable if there exists an algorithm (a Turing machine) that can implement the the function. 

\begin{definition}
A set of natural numbers is called recursively enumerable if there is a computable function f such that for each number $n$, $f(n)$ is defined if and only if n is in the set.
\end{definition}

\noindent Thus a set is recursively enumerable if and only if it is the domain of some computable function. \\

Operations or transformations that are recursively enumerable simply implies that such operations can be implemented as a computer program running on a universal Turing machine.

\subsection{Group-theoretic symmetry}

\begin{definition} Let $X$ be an object in $R^3$. A symmetry axe $l$ of $X$ is a line about which there exists $\theta \in (0, 2\pi)$ such that the object $X$ is rotate by an angle $\theta$ which appears to be identical to $X$. 
\end{definition}

\begin{definition} A symmetry plane $P$ of $X$ is a reflected mirror image of the object $X$ appearing unchanged.
\end{definition}

In group theory, the symmetry group of an object is the group of all transformations under which the object is invariant under composition of the group operations. Here we consider symmetry groups in Euclidean geometry and polyhedral networks.

For example, given an equilateral triangle, the counterclockwise rotation by 120 degrees around the centre leaves the triangle invariant as it maps every vertex and edge of the triangle to another one occupying exactly the same space.

\begin{definition} 
The direct symmetry group of an object $X$, denoted $S_d(X)$, is a group of symmetry of $X$ if only rotation is allowed. 
\end{definition} 

\begin{definition} The full symmetry
group of an object $X$, denoted $S(X)$, is the symmetry group of $X$ of rotations and reflections.
\end{definition} 

A (full) symmetry group is thus a set of symmetry-preserving operations, such as rotations and reflections. Dual polyhedra have the same symmetry group. For example, a tetrahedron has a total of 24 symmetries, that is, $|S(T)| = 24$.


While symmetry groups are continuous or discrete, here we are interested in recursively enumerable discrete symmetry groups and actions such that for every point of the space the set of images of the point under the isometries in the full symmetry group is a recursively enumerable set. 

In what follows we will use this recursively enumerable variation of the group-theoretic characterization of mathematical symmetry, symmetry will be taken thus as a space or graph-theoretic invariant under recursively enumerable defined transformations.

The direct and full symmetry groups of tetrahedra, cubes and octahedra, and dodecahedra and icosahedra are, respectively, $A_4$ and $S_4$, $S_4$ and
$S_4 \times \mathbb{Z}_2$, and $A_5$ and $A_5 \times \mathbb{Z}_2$ suggesting a natural but limited symmetry partial order, the largest the group subindex the more symmetric. However, it is not clear how to compare different types of symmetry.

In the case of the sphere, it is characterized by spherical symmetry, a type of continuous symmetry as it has an infinite number of symmetries (both for rotations and reflections), here we will only require that the scalars and reflections lines involved are recursively enumerable.

Here we will advance a notion of symmetry based both on group theoretic and algorithmic information to find the correspondences between each other. As a result, we will provide a proposal of a total order of symmetry.

\subsection{Information theory}

A random configuration of, let's say a gas in a room, has little symmetry but high entropy, but a specific symmetric configuration will have high entropy because any change will destroy the symmetry towards a more likely disordered configuration. But the extent to which entropy can be used to characterize symmetry is limited to only apparent symmetry, and it is not robust in the face of object description, due to its dependence on probability distributions~\cite{zkpaper}. Entropy measures the uncertainty associated with a random variable, i.e. the expected value of the information in a message (e.g. a string) in bits. Formally, 

\begin{definition}
The Shannon entropy of a string $x$ is defined by:


$$- \sum_{i=1}^n p(x_i) \log_2 p(x_i)$$
\end{definition}

Shannon entropy allows the estimation of the average minimum number of bits needed to encode a string (or object) based on the alphabet size and the occurrence of the alphabet symbolsbased on a probability distribution. Despite its limitations, classical information theory (Shannon entropy) can be consistent with symmetry. One of its main properties is the property of symmetry given by $H(x_1, x_2, \ldots, x_n)=H(x_{\tau(1)}, x_{\tau(2)}, \ldots, x_{\tau(n)})$, where $\tau$ is any permutation from 1 to $n$, and this property also holds for variations of $H$ such as block entropy, where the string bits are taken up by tuples, e.g. bytes. In other words, $H$ remains invariant amidst the reordering of elements. While entropy may look as if it preserves certain properties for symmetrical objects, it also fails to some extent to characterize symmetry. For example, $s_1=0000011111$ and $s_2=1101001011$ have $H(s_1)=H(s_2)$ because there is a permutation $\tau$ that sends $s_1$ onto $s_2$ and vice versa. However, $H$ misses the fact that $s_2$ looks significantly less symmetrical than $s_1$, which has a reflection symmetry at the centre bit. When taking 2-bit elements as units for $H$, hence applying what we will call block entropy denoted by $H_2$ for blocks of size two bits), this is solved and $H_2(s_1) < H_2(s_2)$, but then we will miss possible 2-bit symmetries, and so on. Taking $H_b$ for $b=1, \ldots, n$ where $n=|s_i|$. We will see that for algorithmic complexity, there are nearly similar results, but with far more interesting subtleties.

\subsection{Graph entropy}
\label{entropy}

We will define the Shannon entropy (or simply entropy) of a graph $G$ represented by its adjacency matrix $A(G$) by

\begin{definition}
$H(A(G))=-\sum_{i=1}^n P(A(x_i)) \log_2 P(A(x_i))$
\end{definition}

\noindent where $G$ is the random variable with $n$ possible outcomes (all possible adjacency matrices of size $|V(G)|$). For example, a completely disconnected graph $G$ with all adjacency matrix entries equal to zero has entropy $H(A(G))=0$ because the number of different symbols in the adjacency matrix is 1. However, if a different number of 1s and 0s occur in $A(G)$, then $H(A(G))\neq0$. In general we will use Block Entropy in order to detect more graph regularities (through the adjacency matrix) at a greater resolution. But for Block Entropy there is an extra factor to be taken into account. The adjacency matrix of a graph is not invariant under \textit{graph relabellings}. This means that the correct calculation of the Block Entropy (not relevant for 1-bit Entropy) of a graph has to take into consideration all possible adjacency matrix representations for all possible labellings. Therefore, 

\begin{definition}
The Block Entropy of a graph is given by:

$$
H(G)=\min\{H(A(g_L)) | G_L \in L(G)\}
$$

\noindent where $L(G)$ is the group of all possible labellings of $G$.
\end{definition}

\subsection{Algorithmic complexity}

The algorithmic (Kolmogorov-Chaitin) complexity of a string $x$ is the length of the shortest effective description of $x$. There are several versions of this notion. Here we use mainly the plain complexity, denoted by $C(x)$, and the conditional plain complexity of a string $x$ given a string $y$, denoted by $C(x|y)$, which is the length of the shortest effective description of $x$ given $y$. The formal definitions are as follows. We work over the binary alphabet $\{0, 1\}$. A string is an element of $\{0, 1\}^*$. If $x$ is a string, $|x|$ denotes its length. Let $M$ be a universal Turing machine that takes two input strings and outputs one string. For any strings $x$ and $y$,

\begin{definition}
The algorithmic complexity of $x$ conditioned by $y$ with respect to $M$ is defined as, 
$$C_M (x | y) = \min\{|p| \textit{ such that } M(p, y) = x\}.$$
\end{definition}

We will often drop the subscript $M$ in $C_M (x | y)$ because of the \textit{invariance theorem}, and we will also write $C(x)$ instead of $C(x | \lambda)$ (where $\lambda$ is the empty string). If $n$ is a natural number, $C(n)$ denotes the algorithmic complexity of the binary representation of $n$. Prefix-free complexity $K$ is defined in a similar way, the difference being that the universal machine is required to be prefix-free. That is, only self-delimited programs are valid programs; no program is a prefix of any other, a property necessary to keep $0 < m(s) < 1$ a (semi-) probability measure. 



\subsection{Algorithmic probability}
\label{bdm}

Algorithmic Probability is a seminal concept in the theory of algorithmic information. The algorithmic probability of a string $s$ is a measure that describes the probability that a valid random program $p$ produces the string $s$ when run on a universal Turing machine $U$. In equation form this can be rendered as 

\begin{definition}
$$m(s) = \sum_{p:U(p) = s} 1/2^{|p|}$$
\end{definition}

\noindent That is, the sum over all the programs $p$ for which $U$ outputs $s$ and halts.\\

The Algorithmic Probability~\cite{solomonoff,levin} measure $m(s)$ is related to algorithmic complexity $K(s)$ in that $m(s)$ is at least the maximum term in the summation of programs, given that the shortest program carries the greatest weight in the sum. 

\begin{definition}
The Coding Theorem further establishes the connection between $m(s)$ and $K(s)$ as follows: 

\begin{equation}
\label{codingtheorem}
|-\log_2 m(s) - K(s)| < c
\end{equation}
\end{definition}

\noindent where $c$ is a fixed constant independent of $s$. The Coding Theorem implies that~\cite{cover,calude} one can estimate the algorithmic complexity of a string from its frequency. By rewriting Eq.~\ref{codingtheorem} as: 

\begin{equation}
\label{ctm}
K_m(s)=-\log_2 m(s) + c
\end{equation}

\noindent where $\mathcal{O}(1)$ is a constant. One can see that it is possible to approximate $K$ by approximations to $m$ (such finite approximations have also been explored in~\cite{finite} on integer sequences), with the added advantage that $m(s)$ is more sensitive to small objects~\cite{d4} than the traditional approach to $K$ using lossless compression algorithms, which typically perform poorly for small objects (e.g. small graphs).

As shown in~\cite{zenilgraph}, estimations of algorithmic complexity are able to distinguish complex from random networks (of the same size, or asymptotically growing), which are both in turn distinguished from regular graphs (also of the same size). $K$ calculated by the BDM assigns low algorithmic complexity to regular graphs, medium complexity to complex networks following Watts-Strogatz or Barab\'asi-Albert algorithms, and higher algorithmic complexity to random networks. That random graphs are the most algorithmically complex is clear from a theoretical point of view: nearly all long binary strings are algorithmically random, and so nearly all random unlabelled graphs are algorithmically random~\cite{zenilgraph}.

\subsubsection{The Coding Theorem and Block Decomposition methods}

The \textit{Coding Theorem Method} (CTM)~\cite{d4,d5} is rooted in the relation provided by Algorithmic Probability between frequency of production of a string from a random program and its Kolmogorov complexity (Eq.~\ref{codingtheorem}, also called the algorithmic \textit{Coding theorem}, as distinct from another coding theorem in classical information theory). Essentially it uses the fact that the more frequent a string (or object), the lower its algorithmic complexity; and strings of lower frequency have higher algorithmic complexity.

Here we report results that would not have been possible if they were not specific enough to correctly identify small patterns that represent signatures of the algorithmic content of an object by using CTM and BDM. We show that the AP-based measures either constitute an equivalent or a superior alternative to other more limited measures, such as lossless compression algorithms, widely used as estimators of algorithmic complexity, and to Shannon entropy and related measures that are based on traditional statistics and require that broad assumptions be encoded in their underlying probability distributions.

The Block Decomposition method (BDM) consists in determining the algorithmic complexity of a matrix by quantifying the likelihood that a random Turing machine operating on a 2-dimensional tape (also called a \textit{termite} or \textit{Langton's ant}~\cite{langton}) can generate it and halt. The \textit{Block Decomposition Method} (BDM) decomposes the adjacency matrix of a graph into smaller matrices for which we can numerically calculate its algorithmic probability by running a large set of small 2-dimensional deterministic Turing machines, and therefore its algorithmic complexity upon application of the algorithmic Coding theorem. Then the overall complexity of the original adjacency matrix is the sum of the complexity of its parts, albeit a logarithmic penalization for repetitions, given that $n$ repetitions of the same object only add $\log_2 n$ complexity to its overall complexity, as one can simply describe a repetition in terms of the multiplicity of the first occurrence. The following graph complexity definition will also introduce BDM.

\subsection{The algorithmic complexity of a graph}

We define the algorithmic complexity estimation of a graph as follows: 

\begin{definition}
The Kolmogorov complexity of a graph $G$ is defined as follows:

\begin{equation}
\label{newecaeq}
BDM(g,d) = \sum_{(r_u,n_u)\in A(G)_{d\times d}} \log_2(n_u)+CTM(r_u)
\end{equation}
\end{definition}

\noindent where $K_m(r_u)$ is the approximation of the algorithmic (Kolmogorov-Chaitin) complexity of the subarrays $r_u$ using the algorithmic Coding theorem (Eq.~\ref{ctm}) method that we denote by CTM, $A(G)_{d\times d}$ represents the set with elements $(r_u,n_u)$ obtained when decomposing the adjacency matrix of $G$ into non-overlapping squares of size $d$ by $d$. In each $(r_u,n_u)$ pair, $r_u$ is one such square and $n_u$ is its multiplicity (number of occurrences). From now on $K_{BDM} (g,d=4)$ may be denoted only by $K(G)$, but it should be taken as an approximation to $K(G)$ unless otherwise stated (e.g. when referring to the theoretical true $K(G)$ value).

Considering relabellings, the correct evaluation of the algorithmic complexity of a graph is given by:

\begin{definition}
$$
K(G)=\min\{K(A(G_L)) | G_L \in L(G)\}
$$
\end{definition}

\noindent where $L(G)$ is the group of all possible labellings of $G$.

One contribution of these algorithmic-based measures is that the 2-dimensional versions of both CTM and BDM are native bidimensional measures of complexity and thus do not destroy the 2-dimensional structure of an adjacency matrix.

By making a generalization of the algorithmic Coding theorem using 2-dimensional Turing machines. This makes it possible to define the probability of production of an adjacency matrix as the result of a randomly chosen deterministic 2-dimensional-tape Turing machine without any array transformations to a string, thus making it dependent on yet another mapping between graphs and strings, unlike our approach that natively deals directly with the complexity of the graph adjacency matrix. 

Most algorithms implementing a computable measure of graph complexity are based either on a graph-theoretic/topological feature that is computable or upon Shannon entropy. An example of an improvement on Shannon entropy is the introduction of graph lossless compression such as Graphitour~\cite{graphitour}. A drawback of graph compression is that lossless compression based on popular algorithms such as LZW (Gzip, PNG, Compress), that are traditionally considered to be approximations to algorithmic (Kolmogorov) complexity, are more closely related to Shannon entropy than to algorithmic complexity (which we will denote by $K$). This is because these popular algorithms implement a method that traverses an object looking for trivial repetitions from which a basic grammar is produced based on frequency. 

A major improvement in the means of approximating the algorithmic complexity of strings, graphs and networks, based on the concept of algorithmic probability (AP), offers different and more stable and robust approximations to algorithmic complexity by way of the so-called algorithmic Coding theorem (c.f. below). The method, called the Coding Theorem Method, suffers the same drawback as other approximations of $K$, including lossless compression, related to the additive constant involved in the \textit{Invariance Theorem} as introduced by Kolmogorov, Chaitin and Solomonoff~\cite{kolmo,chaitin,solomonoff}, which guarantees convergence towards $K$ though its rate of convergence is unknown. The chief advantage of the algorithm is, however, that algorithmic probability (AP)~\cite{solomonoff,levin} not only looks for repetitions but for algorithmic causal segments, such as in the deterministic nature of the digits of $\pi$, without the need of distribution assumptions. As with $\pi$, a graph that is produced recursively enumerable will be eventually characterized by algorithmic probability as having low algorithmic complexity, unlike traditional compression algorithms that implement a version of classical block Shannon entropy. In previous work, this kind of recursively enumerable graph~\cite{zkpaper} has been featured, illustrating how inappropriate Shannon entropy can be when there is a need for a universal, unbiased measure where no feature has to be pre-selected.

The method studied and applied here was first defined in~\cite{kolmo2d,bdmpaper} and is in many respects independent of the observer to the greatest possible extent. For example, unlike popular implementations of lossless compression used to approximate algorithmic complexity (such as LZW), the method based on Algorithmic Probability averages over a large number of computer programs found to reproduce the output, thus making the problem of the choice of enumeration less relevant compared to the more arbitrary choice of lossless compression algorithm. The advantage of the algorithmic complexity measure is that when it diverges from algorithmic complexity (because it requires unbounded increasing computational power) it then collapses into Shannon entropy~\cite{bdmpaper}.

We have previously reported connections between algebraic and topological properties using
algorithmic complexity~\cite{zenilgraph}, where we introduced a definition and numerical method for labelled graph complexity; and in applications to the clustering capabilities of network superfamilies in~\cite{zenilkiani}, as well as in applications to biology~\cite{methodszenil}, where we also introduced a generalization of unlabelled graph complexity. Here we carry further these information content approaches for characterizing biological networks and networks in general. We provide theoretical estimations of the error of approximations to the algorithmic complexity of graphs and complex networks, offering both exact and numerical approximations.

The algorithm here considered can deal with a variety of graph types including directed graphs and weighted graphs. The resulting structure could be used for representation and classification as we will see.

\section{Results}

\begin{figure}[h]
  \centering
\includegraphics[width=9.7cm]{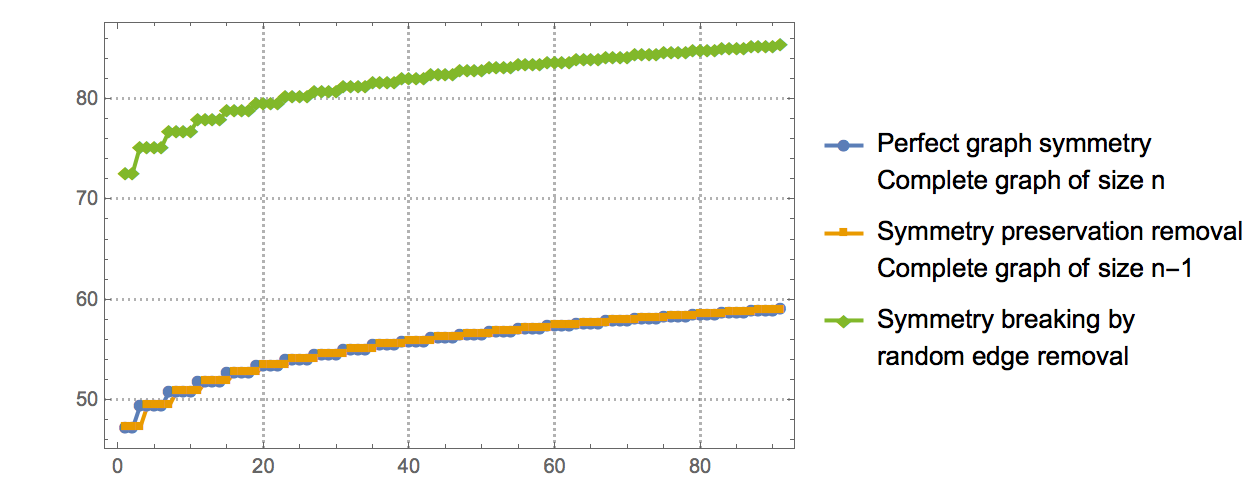}\hspace{.3cm}\includegraphics[width=9.7cm]{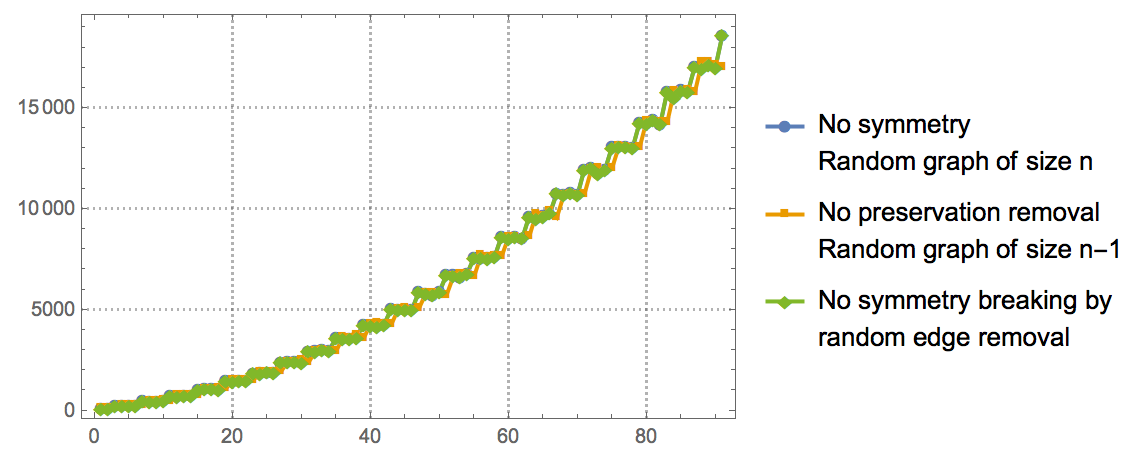}
   \caption{\label{symmetrybreaking}Symmetry breaking. Top: Starting from a growing complete graph (perfect symmetry), removing a node produces another perfectly symmetrical object (another smaller complete graph) hence preserving the symmetry. But as soon as a random edge is deleted from the complete graph, the symmetry is broken and information is generated as quantified by the difference between the original algorithmic content of the complete graph and the mutated one without a single edge. Bottom: In a random world, symmetry breaking does not produce algorithmic information. Entropy and lossless compression both fail to characterize these instances of graph-theoretic symmetry breaking.}
\end{figure}

\subsection{Algorithmic characterization of geometric symmetry}

It is not difficult to identify a group of (recursively enumerable) symmetries as being of low conditional algorithmic (Kolmogorov-Chaitin) complexity because a recursively enumerable transformation requires an encoding of fixed computer program length implementing the recursively enumerable transformation. 

Here we propose an algorithmic information characterization of symmetry:

\begin{definition}
We define a transformation of a recursively enumerable object $s$ to $s^{\prime}$ as a transformation $T$ such that there exists $c=|U(p)=T|$ such that $|K(s^{\prime}) - K(s)| \leq c$
\end{definition}

\noindent where $c$ is the length of a program that implements the transformation $T$ on a (prefix-free) Turing machine running program $p$, and so $K(s^{\prime}) = K(s) + c^{\prime}$ such that $c^{\prime} \leq c$, i.e. $K(s^\prime) \sim K(s)$. The Kolmogorov complexity of the most symmetric geometrical object, the $n$-dimensional hypersphere can be given by: 

\begin{definition}
$$K(c)=\min\{p : p(r,n) = \left\{ x \in \mathbb{Q}^{n} : \|x\| = r\right\}\}$$
\end{definition}

\noindent where $x$ is a computable number. It can be seen that $K$ depends only on the dimension $n$ and the radius $r$.

Let $T(s)$ be the recursively enumerable symmetry group for object $s$. It follows that if there is a recursively enumerable function $t \in T$ such that $s^\prime=t(s)$ with $t$ and a recursively enumerable function $t^{-1}$ mapping $s^\prime$ to $s$, then $|K(s) - K(s^\prime) |< c$. In other words, $K$ is invariant.

For example, a tetrahedron $s$ can be placed in 12 distinct positions by rotation alone $s_i$ for $i \in \{1, \ldots, 12\}$. The 12 rotations form the rotation (symmetry) group. Let $t$ be one of these rotations. Without loss of generality, if $t$ is recursively enumerable, let $p(t)$ be the program implementing $t$ such that $s_i=t(s)$, then $|K(s) - K(s^\prime)| < |p(t)|$.

Just as in Euclidean geometry, algorithmic complexity remains invariant under symmetric transformations in any space using the same argument, with the only requirement that $t \in T$ is a recursively enumerable transformation acting in a recursively enumerable space.

\begin{figure}[htp!]
  \centering
  (a)\\
  
  \includegraphics[width=12cm]{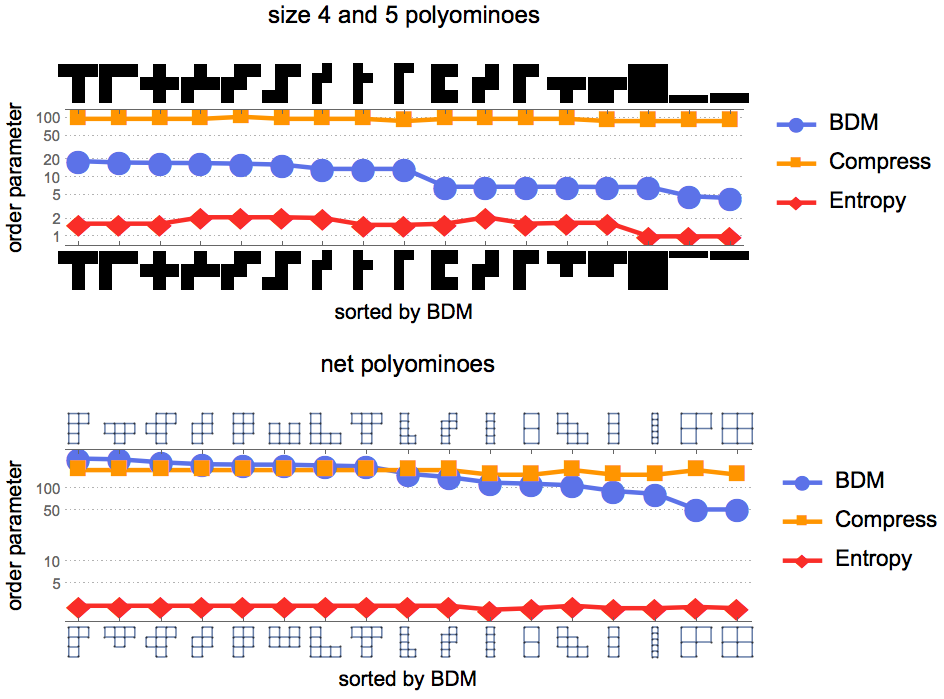}\\
  \bigskip
  \medskip
  
 (b)\hspace{5cm}(c)\hspace{2cm}\\
  
\includegraphics[width=7.8cm]{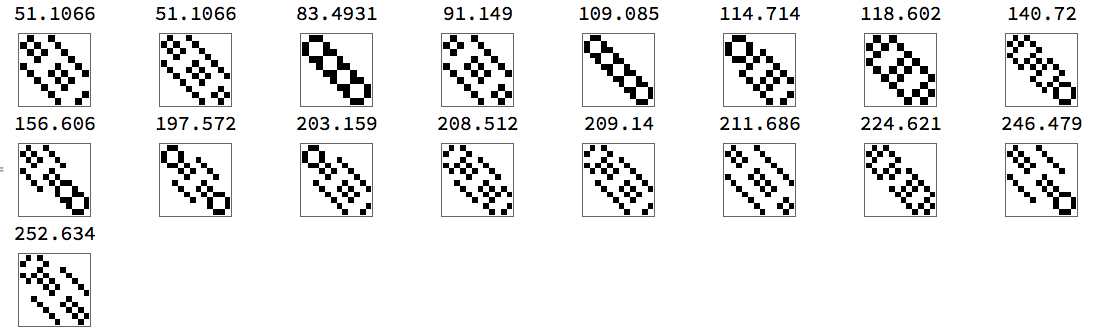}\hspace{.5cm}\includegraphics[width=3.8cm]{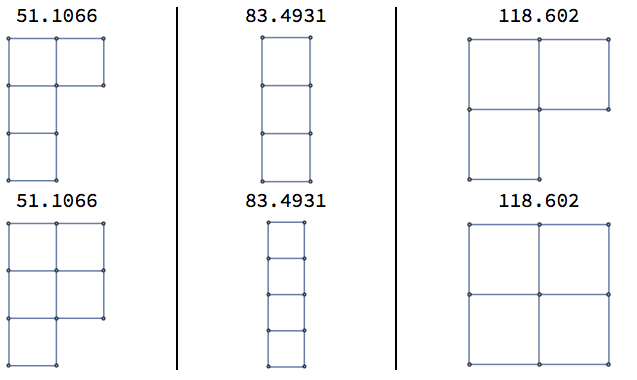}
\caption{Adjacency matrices of net-form representation of simply connected free polyominoes (without holes). They look very different from their cell representations, yet the AP-based measure produces a robust classification independent of object description. Spearman rank correlation values between the same polyominoes represented by simple arrays (a) or as graphs, and thus as different arrays as represented by their adjacency matrices (b) were:
$\rho = 0.99, p = 3.216\times 10^{-14}$ for AP-based ranking; $\rho =0.44, p =0.0779$ for Compress; and $\rho = 0.227, p = 0.38$ and thus only statistically significant for the AP-based ranking. The AP-based measure also assigns the same complexity to qualitatively similar polyominoes (c), something that neither compression nor entropy was able to do consistently, because of under- or over-sensitivity. (b) and (d) display their AP-estimated value on top of each element. Some functions from a library from Eric Weinstein's MathWorld were adapted for our purposes.~\cite{mathworld}} 
  \label{polyos}
\end{figure}

\subsection{Algorithmic symmetry ranking}

The above suggests a symmetry ranking of objects subject to symmetry comparison that can be achieved numerically as demonstrated in Fig.~\label{polyhedra1}. The algorithmic characterization of symmetry is a particular case of the more general case of algorithmic complexity alone and is thus rather an application of algorithmic complexity to symmetry with the provisions of recursivity/computability required.

Fig.~\ref{symmetrybreaking} illustrates the way in which estimations of the algorithmic information of graphs capture symmetry breaking, thereby demonstrating the ability of these methods to characterize it. Fig.~\ref{symmetrybreaking}(top) shows how removing edges from a complete graph drives the estimations much higher  while removing nodes that preserve the symmetries of the complete graph (by producing another complete graph) and thus remaining stable for graphs of growing size. In contrast, Fig.~\ref{symmetrybreaking}(bottom) illustrates that random graphs are immune to symmetry breaking, being deprived of symmetries to begin with, and both node and edge removal have the same effect.

Fig.~\ref{symmetrybreaking}(top) illustrates how symmetry breaking generates information. The results show how to detect and produce information content starting from a highly symmetrical object (a complete graph). Similar results had been reported in~\cite{zenilgraph}. It also suggests how growing a symmetrical object converges and cannot be the source of new information if unbroken; only symmetry breaking can generate differences that produce information when moving a symmetrical object from simplicity to complexity. Likewise, information from breaking a random object does not produce information that could not already be generated by the underlying source of random graphs. 

\begin{figure}[h]
  \centering
\includegraphics[width=10cm]{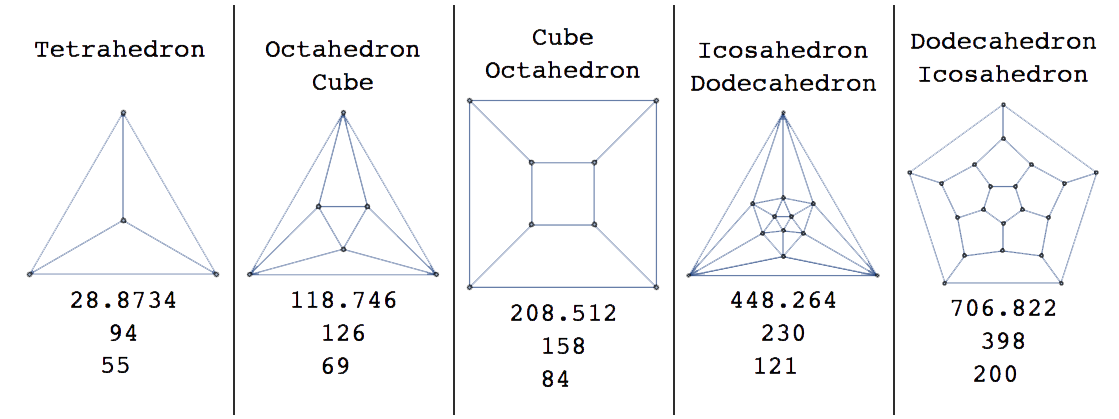}\\ 
\bigskip

\includegraphics[width=6.8cm]{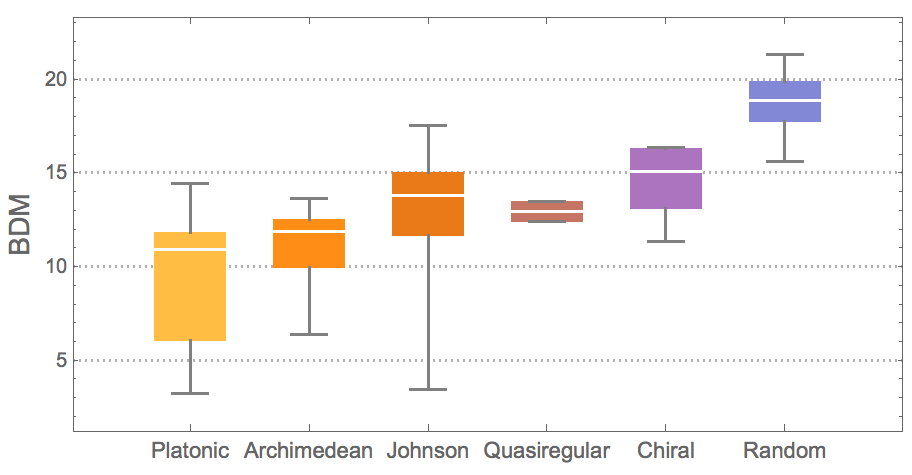}\hspace{.7cm} \includegraphics[width=4.4cm]{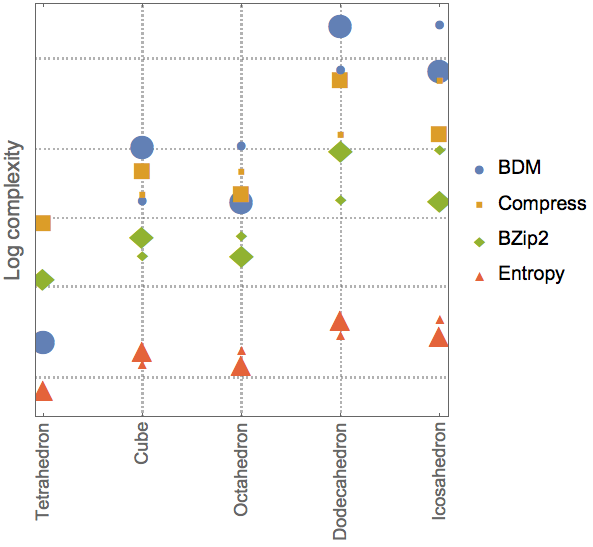}
   \caption{Top: Platonic solid networks with order parameter values (BDM, Entropy and lossless compression by Compress). Bottom left: Different polyhedra sorted by algorithmic symmetry (BDM). Bottom right: Platonic graphs and their duals classified by order parameters, duals are in same colour. Numerical approximations of graphs and duals have similar values, as theoretically expected, given that there is an algorithm of fixed size that sends a graph to its unique dual and back.}
\label{polyhedra1}
\end{figure}

\begin{figure}[h]
  \centering
\includegraphics[width=5.7cm]{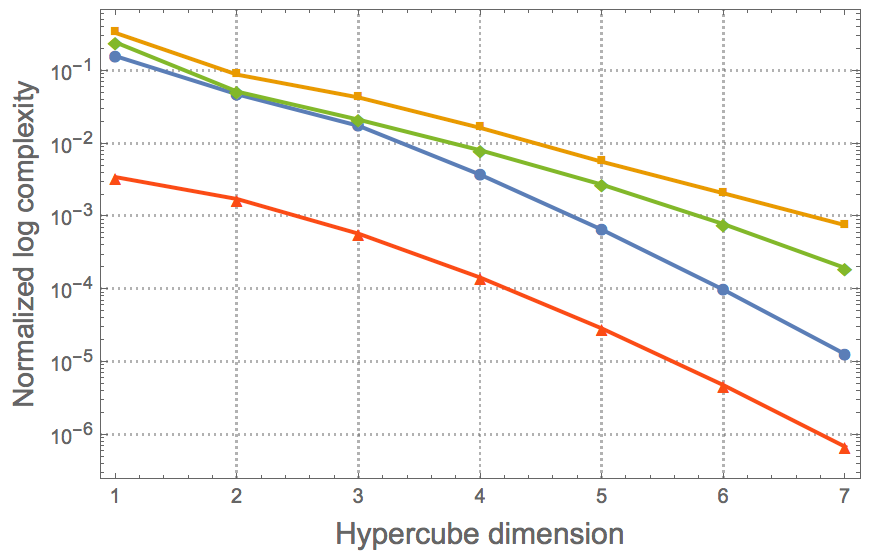}\hspace{.6cm} \includegraphics[width=5.7cm]{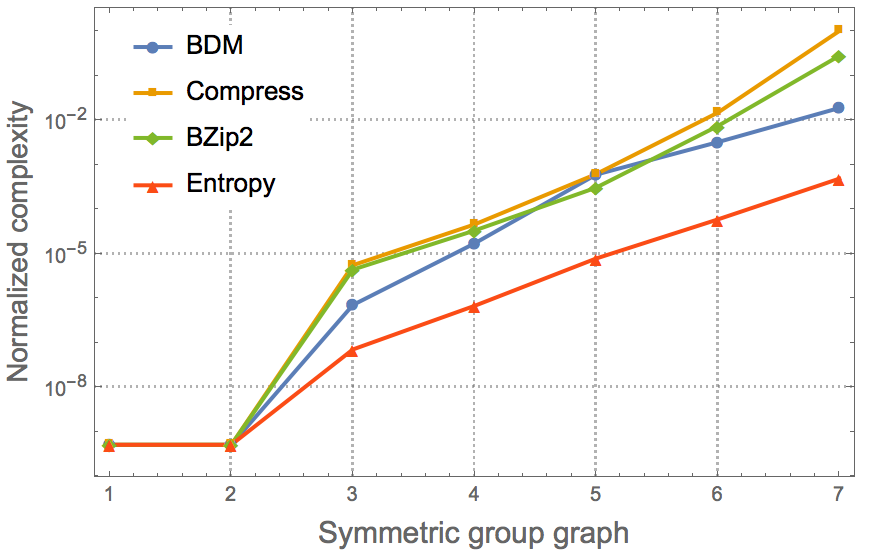}
   \caption{Left: Numerical approximation of algoritmic complexity by BDM over a polytope (hypercube) of growing dimensions. Graphs of different symmetric groups have different growing numerical algorithmic complexity values.}
\label{dimension}
\end{figure}

\subsection{Classification of polyominoes}

When exploring free polyominoes (equivalent shapes under the symmetry group are considered the same) of size 4 and 5 non-zero (black) cells, we found that the classification by AP-based measures can yield more robust characterizations in the sense that they are less variant amidst changes in the description of the same object. In Fig~\ref{polyos}, we classified the same set of polyominoes by their geometric versus topological representations. In the case of the geometric representation, we considered a binary array of black and white cells of length $n \times m$ such that each row $1\ldots n$ and column $1\ldots m$ contains at least a non-zero element. The topological representation is a network whose vertices represent the corners of the geometrical representation and whose edges are cell boundaries, of which at least one is shared with another cell. Spearman ranking correlation values are given in Fig~\ref{polyos}, establishing the ability of the AP-measure to deal with small patterns in a robust but sensitive fashion, unlike the two other methods we compared with (a lossless compression algorithm and Shannon entropy). The compression algorithm used is Compress, based on a Unix shell compression program, itself based on the LZW compression algorithm. LZW is the most common lossless compression algorithm behind, e.g., png, zip, gzip. Shannon entropy is applied to the arrays, with the space of all possible uniformly distributed binary arrays of the size of either the bitmap or the adjacency matrix of the network representation as probability space.

\subsection{Polytope profiling} 

Regular polyhedra or so-called Platonic solid networks are of low complexity compared 
to the sphere, for which we showed that its complexity grows by $\log_2 n \log_2 r$ the 
$r$ is the radius and $n$ the dimension of a hypershpere.

Polygonal approximations to sphere rendering depend only on the number of polygons used and therefore will tend to be of greater complexity than, say, platonic solids. Platonic solids are rough approximations of spheres with low polygon face count, and hence are lower bounds of any other sphere approximation.

Unlike the Platonic solids, that are all formed by a single polygon, each Archimedean polyhedron is formed by two different polygons and they are thus expected to be slightly more complex, as numerically shown in Fig.~\ref{polyhedra1}. It shows the expected agreement between the classical information and algorithmic complexity of objects of similar complexity, in particular graphs and their duals, where vertices are exchanged for edges and edges for vertices from the original graphs. Fig.~\ref{dimension} reports the change of complexity as a function of dimension and symmetric group over a hypercube of growing dimension and the set of all graphs grouped by symmetric group extracted from GraphData[] in the Wolfram Language.

\section{Conclusions}

The complexity of graphs has historically been based on graph-theoretic, and more recently, Entropy-based indices. While some of them may continue to be of interest in approaches to molecular similarity (such as QSPR and QSAR), here we have instead explored more universal approaches to the problem of feature-free approximations of the symmetry and complexity of a graph or network. An essential ingredient for a complexity measure is that it integrates all its properties, something that only a noncomputable approximating measure such as CTM/BDM can achieve. We have shown suggestive connections between algorithmic information and symmetry in the context of geometric and graph objects. We have demonstrated that we can properly characterize objects and that their characterization is robust when taking them as geometric or as geometry-based networks in cases of study includong polyominoes and polytopes going beyond group-theoretic characterizations and achieving better results than those achieved by other means such as those based by lossless compression and Shannon entropy.


\section*{Acknowledgements}

H.Z. acknowledges the support of the Swedish Research Council (Vetenskapsr\r{a}det) grant No. 2015-05299.

\newpage

\section*{Appendix}

\begin{figure}[h]
  \centering
\includegraphics[width=6.5cm]{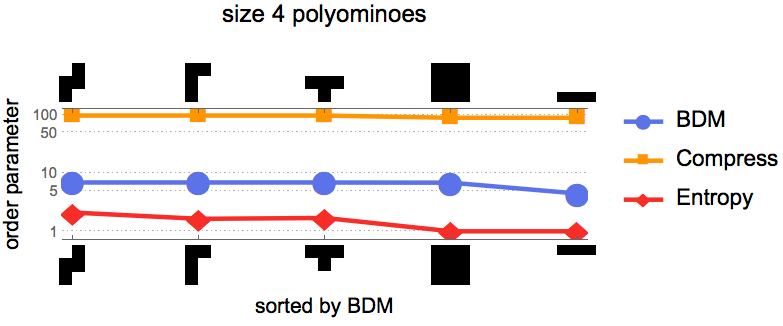}\\
\bigskip

  \includegraphics[width=8.5cm]{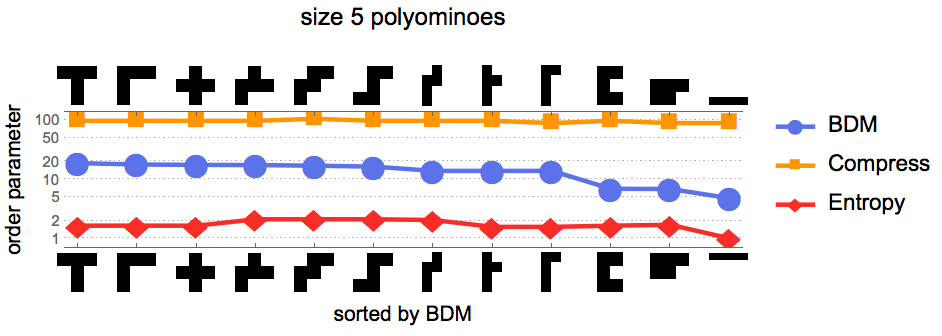}\\
  \bigskip
  
  \includegraphics[width=12cm]{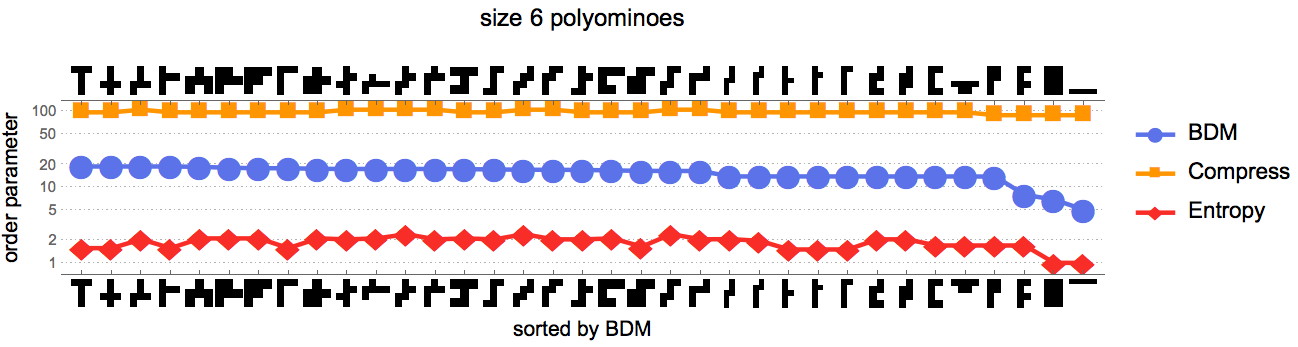}
  \caption{Another illustration of the distribution of polyominoes according to different order parameters. Entropy is sometimes too sensitive or insensitive, lossless compression (Compress) is always too insensitive and the AP-based measure BDM appears to be robust, assigning similar complexity and respecting relative order as shown in Fig.~\ref{polyos}.} 
  \label{polyos2}
\end{figure}


\begin{thebibliography}{100}

\bibitem{bottcher} T. B{\"o}ttcher, An Additive Definition of Molecular Complexity, \textit{J. Chem. Inf. Model.}, 2016, 56 (3), pp 462–470

\bibitem{graphitour} L. Peshkin, Structure induction by lossless graph compression, Proc. 2007 Data Compression Conference, editors: J.A Storer and M. Cohn, pp 53—62, 2007




















\bibitem{gutman} I. Gutman, The energy of a graph: old and new results, In A. Betten, A. Kohner, R. Laue, and A. Wassermann, eds. \emph{Algebraic Combinatorics and Applications}, Springer, Berlin, 196--211, 2001.

\bibitem{fowler} P.W. Fowler, Energies of graphs and molecules, In \emph{Computational Methods in Modern science and Engineering}, vol 2, parts A and B, Corfu, Greece, 517--520, 2007.














\bibitem{bonchev2} D. Bonchev, Overall Connectivity and Topological Complexity: A New Tool for QSPR/QSAR. In J. Devillers, A.T. Balaban, (Eds) \textit{Topological Indices and Related Descriptors in QSAR and QSPR}, Gordon \& Breach: Langhorne, PA, pp 361-401, 1999.









\bibitem{calude} C.S. Calude, \emph{Information and Randomness: An Algorithmic Perspective}, EATCS Series, 2nd. edition, 2010, Springer.

\bibitem{chaitin} G.J. Chaitin. On the length of programs for computing finite binary sequences \emph{Journal of the ACM}, 13(4):547--569, 1966.

\bibitem{cover} T.M. Cover and J.A. Thomas, \emph{Elements of Information Theory}, 2nd Edition, Wiley-Blackwell, 2009.

\bibitem{d4} J.-P. Delahaye and H. Zenil, Numerical Evaluation of the Complexity of Short Strings: A Glance Into the Innermost Structure of Algorithmic Randomness, \emph{Applied Mathematics and Computation} 219, pp. 63--77, 2012.

\bibitem{kolmo} A. N. Kolmogorov. Three approaches to the quantitative definition of information, \emph{Problems of Information and Transmission}, 1(1):1--7, 1965.
\bibitem{levin} L. Levin, Laws of information conservation (non-growth) and aspects of the foundation of probability theory, \emph{Problems in Form. Transmission} 10. 206--210, 1974.
\bibitem{langton} C.G. Langton, Studying artificial life with cellular automata, \emph{Physica D: Nonlinear Phenomena 22} (1--3): 120--149, 1986.




\bibitem{lin} S-K. Lin, Correlation of Entropy with Similarity and Symmetry, \textit{J. Chem. Inf. Comput. Sci.}, Vol. 36, No. 3, 1996.



\bibitem{finite} F. Soler-Toscano, H. Zenil, A Computable Measure of Algorithmic Probability by Finite Approximations with an Application to Integer Sequences, Complexity  vol. 2017, 2017.




\bibitem{solomonoff} R.J. Solomonoff, A formal theory of inductive inference: Parts 1 and 2. \emph{Information and Control}, 7:1--22 and 224--254, 1964.


\bibitem{kolmo2d} H. Zenil, F. Soler-Toscano, J.-P. Delahaye and N. Gauvrit, \emph{Two-Dimensional Kolmogorov Complexity and Validation of the Coding Theorem Method by Compressibility}, 2013.

\bibitem{zenilkiani} H. Zenil, N.A. Kiani and J. Tegn\'er, Algorithmic complexity of motifs clusters superfamilies of networks, Proceedings of the \emph{IEEE International Conference on Bioinformatics and Biomedicine,} Shanghai, China, 2013.



\bibitem{d5} F. Soler-Toscano, H. Zenil, J.-P. Delahaye and N. Gauvrit, ``Calculating Kolmogorov Complexity from the Output Frequency Distributions of Small Turing Machines'', \textit{PLoS ONE} 9(5): e96223.
\bibitem{mathworld} E.W. Weisstein, ``Polyomino." From MathWorld--A Wolfram Web Resource. \url{http://mathworld.wolfram.com/Polyomino.html}
 \bibitem{zenilgraph} H. Zenil, F. Soler-Toscano, K. Dingle and A. Louis, Correlation of automorphism group size and topological properties with program-size complexity evaluations of graphs and complex networks, \textit{Physica A: Statistical Mechanics and its Applications}, Volume 404, pp. 341--358, 2014.
 \bibitem{toxicologypaper} N.A. Kiani, M. Shang, H. Zenil and J. Tegn\'er, Predictive Systems Toxicology. In Orazio Nicolotti (ed.), \textit{Computational Toxicology - Methods and Protocols, Methods in Molecular Biology}, Springer, 2017 (in press).

\bibitem{chemicalnetworks} H. Zenil, Narsis A. Kiani, M-M. Shang, J. Tegn\'er, Algorithmic Complexity and Reprogrammability of Chemical Structure Networks, arXiV preprint.

 \bibitem{methodszenil} H. Zenil, N.A. Kiani and J. Tegn\'er, Methods of Information Theory and Algorithmic Complexity for Network Biology, \textit{Seminars in Cell and Developmental Biology},  vol. 51, pp. 32-43, 2016.
\bibitem{bdmpaper} H. Zenil, F. Soler-Toscano, N.A. Kiani, S. Hern\'andez-Orozco, A. Rueda-Toicen, A Decomposition Method for Global Evaluation of Shannon Entropy and Local Estimations of Algorithmic Complexity, arXiv:1609.00110 [cs.IT].
\bibitem{zkpaper} H. Zenil, N.A. Kiani and Jesper Tegn\'er, Low Algorithmic Complexity Entropy-deceiving Graphs, \textit{Phys Rev E.} 96, 012308, 2017..

 
\end{thebibliography}
\end{document}